\documentclass[5p]{elsarticle}

\usepackage{lineno,hyperref}
\usepackage{dblfloatfix} 
\usepackage{float}
\usepackage{graphicx}
\usepackage{soul} 
\usepackage{xcolor} 
\usepackage{caption}
\usepackage{amssymb}
\usepackage{amsmath}
\usepackage{bm}
\usepackage[normalem]{ulem}
\usepackage{upgreek}
\usepackage[detect-all]{siunitx}
\usepackage{nicefrac}

\modulolinenumbers[10]

\journal{ArXiv}

\begin{document}


\begin{frontmatter}

\title{Advancing in situ hydrogen embrittlement studies through an integrated charging cell for SEM micromechanical testing}


\author[myaddress1]{Lavakumar Bathini\corref{mycorrespondingauthor1}}\cortext[mycorrespondingauthor1]{Corresponding author} \ead{lavakumar-bathini@cnrs.fr}
\author[myaddress1]{Guillaume Kermouche}
\author[myaddress1]{Sergio Sao-Joao}
\author[myaddress1]{Frédéric Christien}

\author[myaddress1]{Szilvia Kal\'{a}cska\corref{mycorrespondingauthor2}}\cortext[mycorrespondingauthor2]{Corresponding author} \ead{szilvia.kalacska@cnrs.fr}

\address[myaddress1]{Mines Saint-Etienne, Univ Lyon, CNRS, UMR 5307 LGF, Centre SMS, 158 cours Fauriel 42023 Saint-Étienne, France}

\begin{abstract}
A comprehensive understanding of hydrogen-deformation interactions at the microscale is essential for revealing hydrogen embrittlement mechanisms. In situ micromechanics with simultaneous hydrogen (H-) charging has therefore gained traction in recent times. In the present study, we aim to address the drawbacks of current in situ H-charging setups by developing a more robust 3-electrode-based back-side charging system for a scanning electron microscope to perform various micromechanical tests. The development of the novel setup is discussed and demonstrated through micropillar compression of an Fe-25Cr single crystal (110) during H-charging. H has increased the yield strength and the apparent strain-hardening rate. H activates multiple slip systems and enhances dislocation density and entanglement, leading to pronounced forest hardening as revealed by electron microscopy. Estimation of activation volume from strain-rate jump tests indicates that the deformation is controlled by the solute drag effect on kink mobility and dislocation forest hardening.

\end{abstract}


\end{frontmatter}

\section{Introduction}

Hydrogen embrittlement (HE) is a well-known material degradation phenomenon arising from a complex interplay among hydrogen (H), the microstructure, and mechanical loading. Understanding how hydrogen influences deformation mechanisms is crucial for the design of HE-resistant materials and, therefore, the growth of the “hydrogen economy” \cite{Yu.2024}. Conventional macroscale HE studies have primarily focused on assessing changes in bulk mechanical properties and interpreting the underlying mechanisms through post-mortem analysis. However, due to its small atomic size and high diffusivity, the interaction of H with material defects (vacancies, dislocations, grain/phase boundaries, etc.) and microstructural features (individual grains/phases, precipitates, etc.) spans multiple length scales. These interactions fundamentally govern deformation and subsequent failure processes \cite{Duarte.2021}. Therefore, developing a mechanistic understanding of HE requires resolving the local hydrogen-microstructure/defect interactions, which cannot be adequately captured through conventional macroscale testing \cite{Duarte.2021,Barnoush.2006,Barnoush.2008,Kim.2019}. Hence, investigating hydrogen-deformation interactions at the sub-micron scale under well-defined mechanical and microstructural boundary conditions is essential \cite{Kim.2019}. The key challenge with such an approach is ensuring the presence of H in the probing volume. Since H outgasses rapidly in most of the materials, studying HE using micromechanical testing mandates in situ H-charging \cite{Barnoush.2019,Sun.2021}. Therefore, over the past two decades, in situ nanoindentation \cite{Duarte.2021,Barnoush.2008,Rao.2023,Barnoush.2010,Zeiler.2024,Rao.2023T}, micropillar compression \cite{Lu.2023,Zhu.2025,Kim.2020,Lu.2021,Kapci.2025,Kheradmand.2012,Barnoush.2010.2}, and microcantilever bending \cite{Deng.2018,Asadipoor.2022,Deng.2017} have emerged as powerful techniques for quantifying the effect of H on local mechanical properties \cite{Massone.2022}.

\begin{figure*}[!hb]
    \centering
    \includegraphics[width=0.8\textwidth]{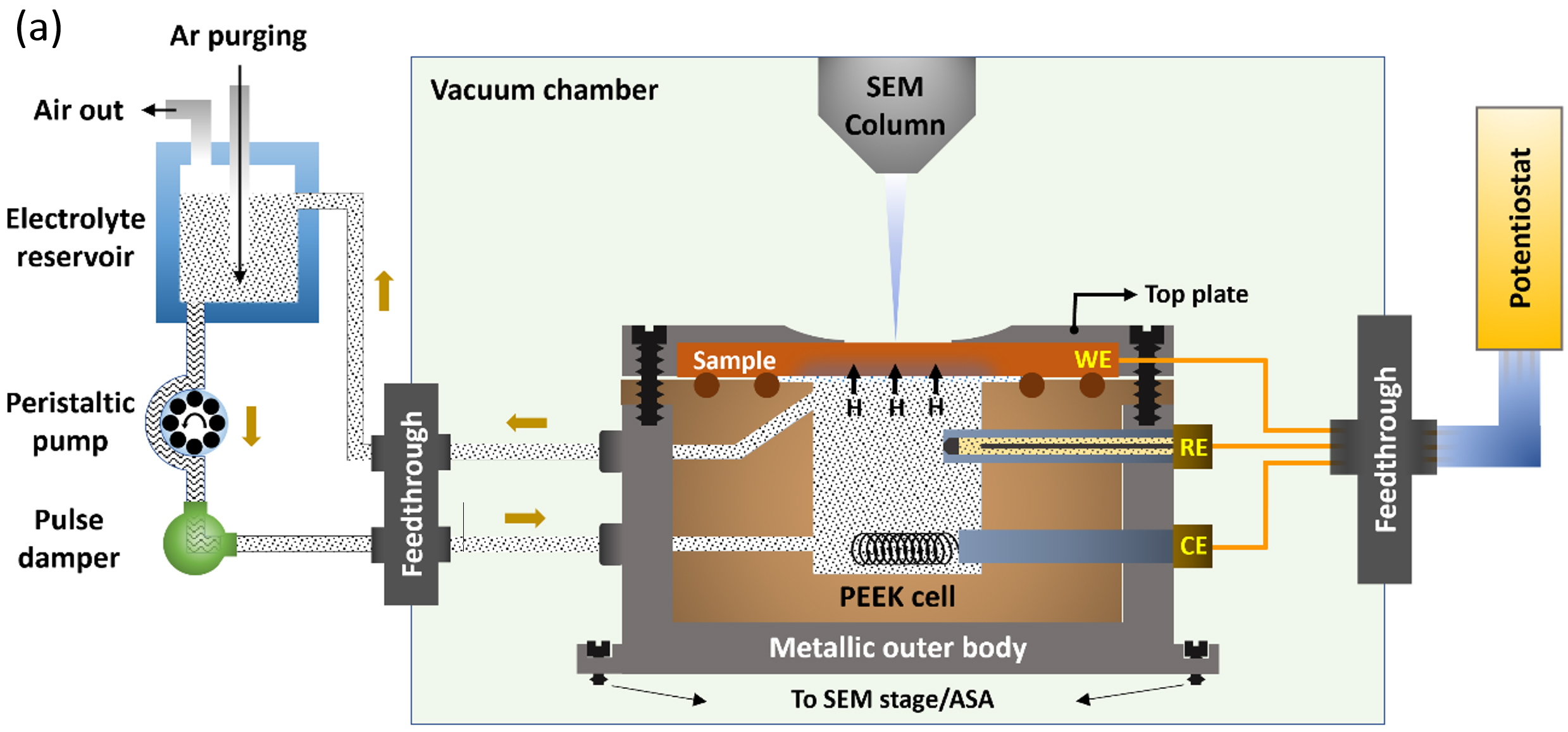}
    \includegraphics[width=0.8\textwidth]{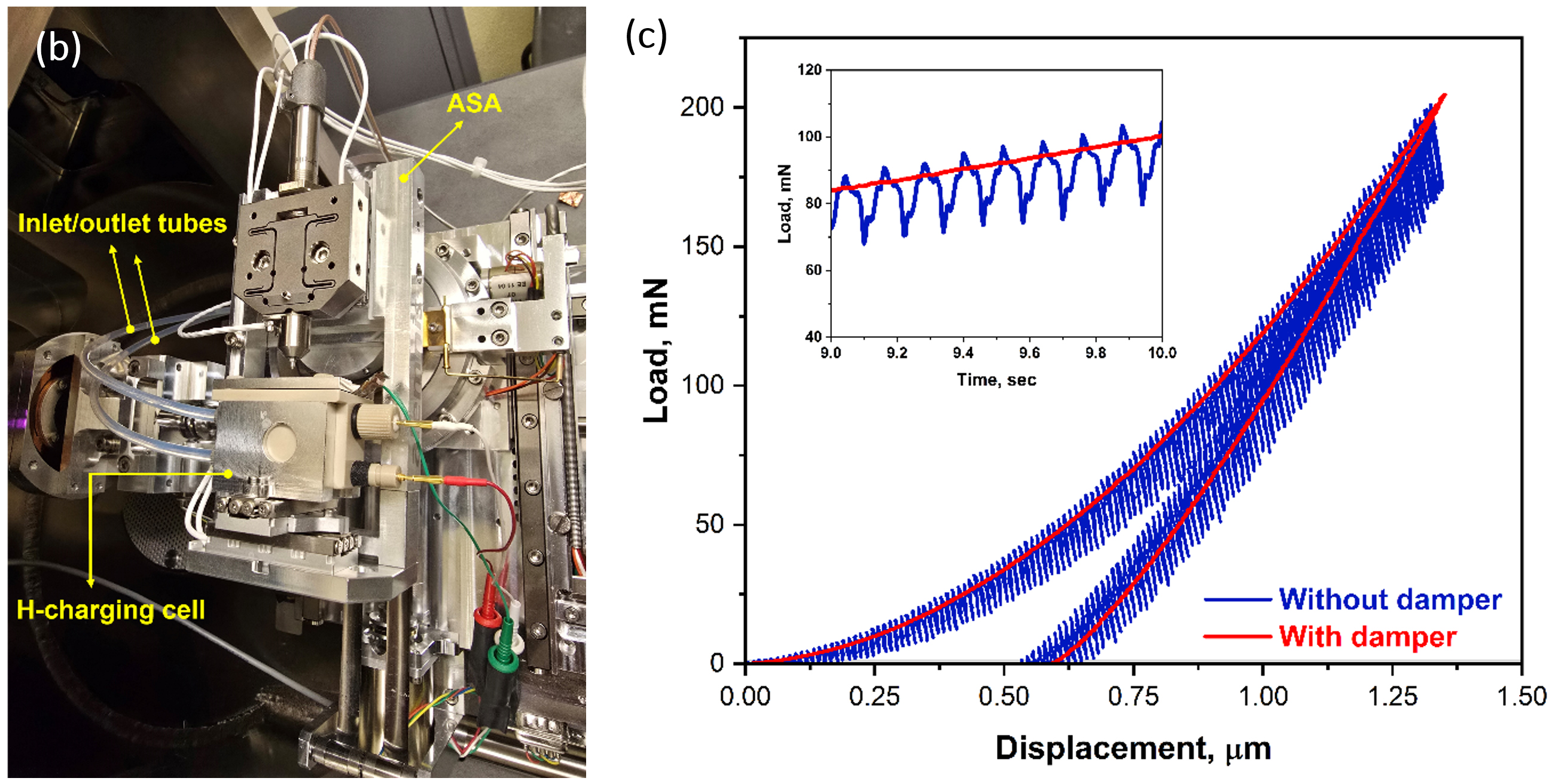}
    \caption{(a) Cross-sectional view of the in-situ H-charging setup for SEM, (b) H-cell along with ASA mounted on the SEM stage, and (c) nanoindentation curves of fused silica during electrolyte circulation obtained with and without the damper.
    \label{fig:asa}}
\end{figure*}

While the in situ electrochemical nanoindentation technique has provided valuable insights into fundamental mechanisms \cite{Barnoush.2010}, there is a growing concern regarding the deleterious effects of the charging method on the surface integrity (due to front-side charging) and thus the validity of the measured properties. For example, Duarte et al. have observed contradictory nanoindentation results with front-side and back-side H-charging methods and attributed it to surface degradation \cite{Duarte.2021}. Similarly, surface damage is evident in on the micropillars \cite{Lu.2023,Kapci.2025} and microcantilevers \cite{Deconinck.2026} tested with front-side charging. Such artefacts can dramatically influence the mechanical properties \cite{Robertson.2001},  especially at the small scale \cite{Barnoush.2019} and also complicate the post mortem analysis \cite{Lu.2023}. Although an electrolyte to preserve the integrity of the charging surface has been proposed \cite{Hajilou.2018}, this can only work for a limited number of materials. For instance, Lu et al. have carried out micropillar compression of TWIP steel [11] and alloy 725 \cite{Lu.2021} using the front-side charging method. In the former case, slip traces could not be distinguished due to corrosion, whereas in the latter case, the pillar surface remained intact. Since these surface features are crucial to elucidate the underlying mechanisms, it is necessary to avoid this problem. Also, unwanted phase transformations on the micropillar surface were observed during front-side charging, complicating the mechanical behaviour \cite{Kapci.2025}. Furthermore, it is known that these charging-related artefacts can play a definitive role in the hardening/softening dichotomy observed in micromechanical properties. For example, Barnoush et al. \cite{Barnoush.2010.2} have observed softening in FeAl micropillars while the same group has reported hardening in Ni micropillars due to H \cite{Kheradmand.2012} despite the reduction in pop-in load in both cases \cite{Barnoush.2010}. Similar inconsistencies can be observed in other studies \cite{Kim.2020,Luo.2026}. In addition, since the probing surface is covered with electrolyte, the front-side charging method does not allow simultaneous imaging during micromechanical testing, preventing accurate assessment of deformation modes, slip activity, and microstructure evolution \cite{Kim.2019}. To circumvent these issues, Kim and Tasan have developed a 2-electrode-based in situ electrochemical back-side H-charging setup compatible with a scanning electron microscope (SEM) \cite{Kim.2019}. However, the confined, narrow geometry of the electrolyte chamber in such a setup can result in rapid hydrogen bubble accumulation at the charging surface, leading to unstable and spatially non-uniform charging conditions. Reliable H-charging, therefore, requires continuous electrochemical monitoring and control, necessitating a reference electrode. Furthermore, electrolyte circulation can induce flow-driven oscillations of the thin samples and affect the micromechanical data \cite{Zeiler.2024}. 

\begin{figure*}[!hb]
    \centering
    \includegraphics[width=0.99\textwidth]{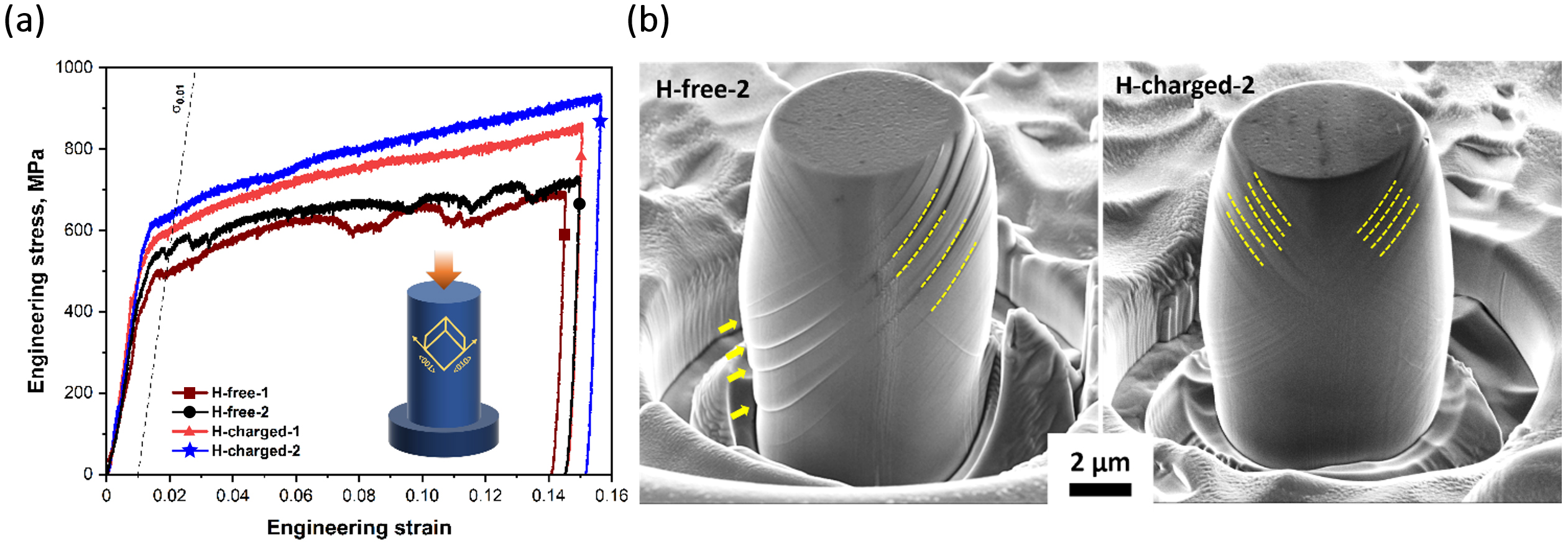}
    \caption{(a) Engineering stress-strain curves of Fe-25Cr micropillars compressed inside the SEM, and (b) post mortem secondary electron images of the deformed pillars.
    \label{fig:pillars}}
\end{figure*}

In the current work, we addressed the aforementioned drawbacks and developed a compact 3-electrode-based H-charging cell compatible with high-vacuum SEM and an in situ micromechanical testing device, demonstrated here using the Alemnis Standard Assembly (ASA). Since this platform employs highly sensitive piezo-driven positioning stages, the cell was designed to minimise size and weight. Figure \ref{fig:asa}(a) shows the cross-sectional schematic of the H-cell, consisting of a metallic support body and an inner PEEK chamber containing the electrolyte. PEEK was selected for its chemical inertness, low water absorption, mechanical strength, machinability, and vacuum compatibility. A disc-shaped thin sample is clamped to the chamber with a top plate, while O-rings provide a vacuum-tight seal to prevent electrolyte leakage. The exposed charging area on the sample's back side is approximately 12 mm in diameter. While the sample serves as the working electrode (WE), a platinum wire acts as the counter electrode (CE). A customised miniature Ag/AgCl reference electrode (RE) is positioned near the sample surface. All electrodes are connected to an external potentiostat (Gamry Reference 620). Electrolyte is continuously circulated through vacuum-compatible PTFE tubing using a peristaltic pump (Innofluid), while a pulse damper installed upstream reduces flow-induced vibrations during in situ testing. Figure \ref{fig:asa}(b) shows the H-cell mounted onto the indenter placed on the SEM stage. The effectiveness of the damping system was evaluated through nanoindentation experiments on fused silica with continuous electrolyte circulation. Figure \ref{fig:asa}(c) compares the load-displacement curves obtained with and without the damper. The integration of the pulse damper significantly suppresses the oscillations originating from pulsations generated by the peristaltic pump (see insert in Figure \ref{fig:asa}(c)), resulting in substantially smoother load–displacement curves. The set\-up thus enables micromechanical testing during controlled active hydrogen charging with simultaneous imaging, allowing direct observation of hydrogen -- microstructure -- deformation interactions in real time. The capabilities of the developed setup are demonstrated through in situ micropillar compression experiments on an Fe-25Cr single crystal. Sample details are shared in the Supplementary Materials.

\begin{figure*}[!hb]
    \centering
    \includegraphics[width=0.68\textwidth]{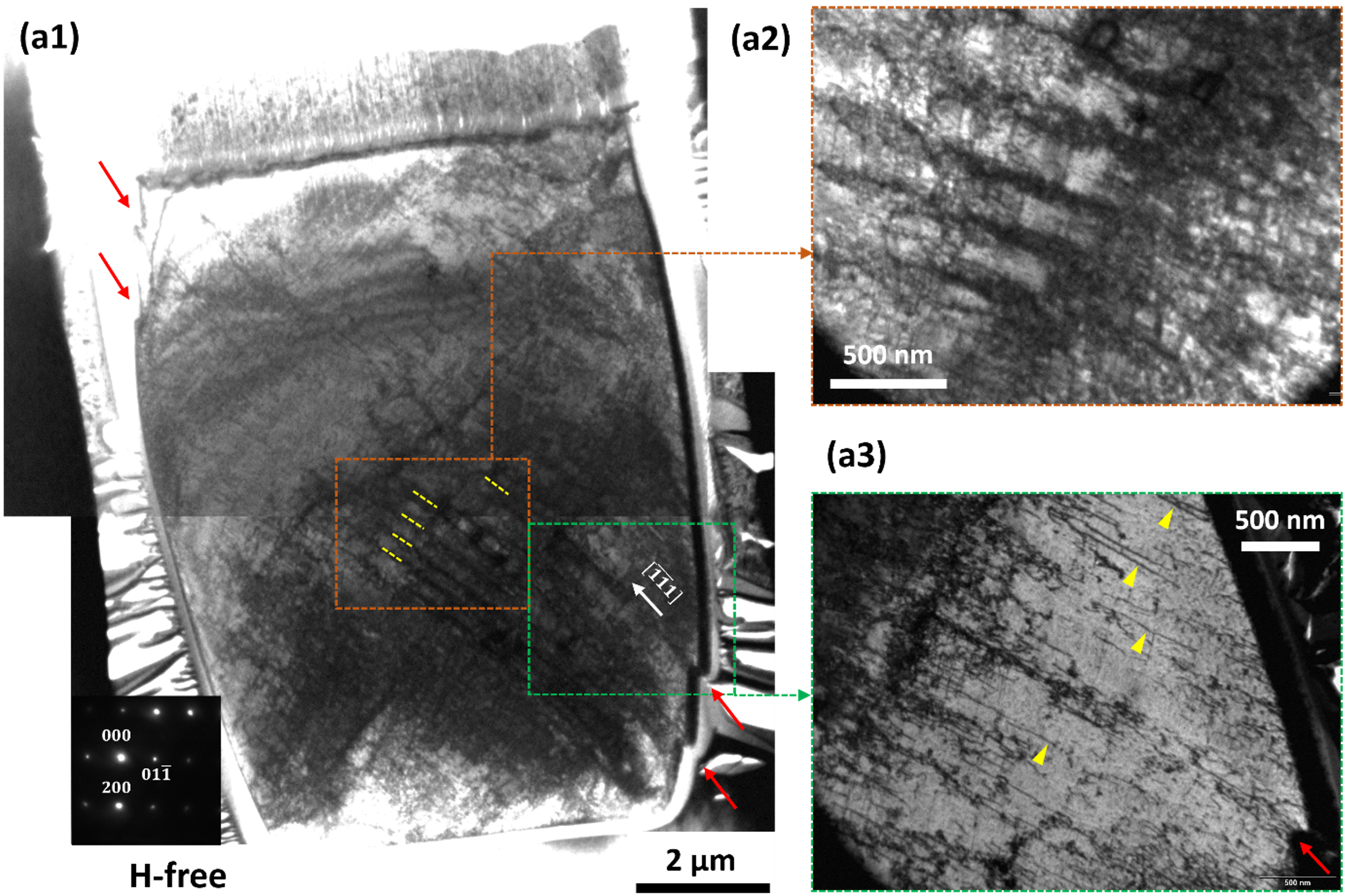}
    \includegraphics[width=0.68\textwidth]{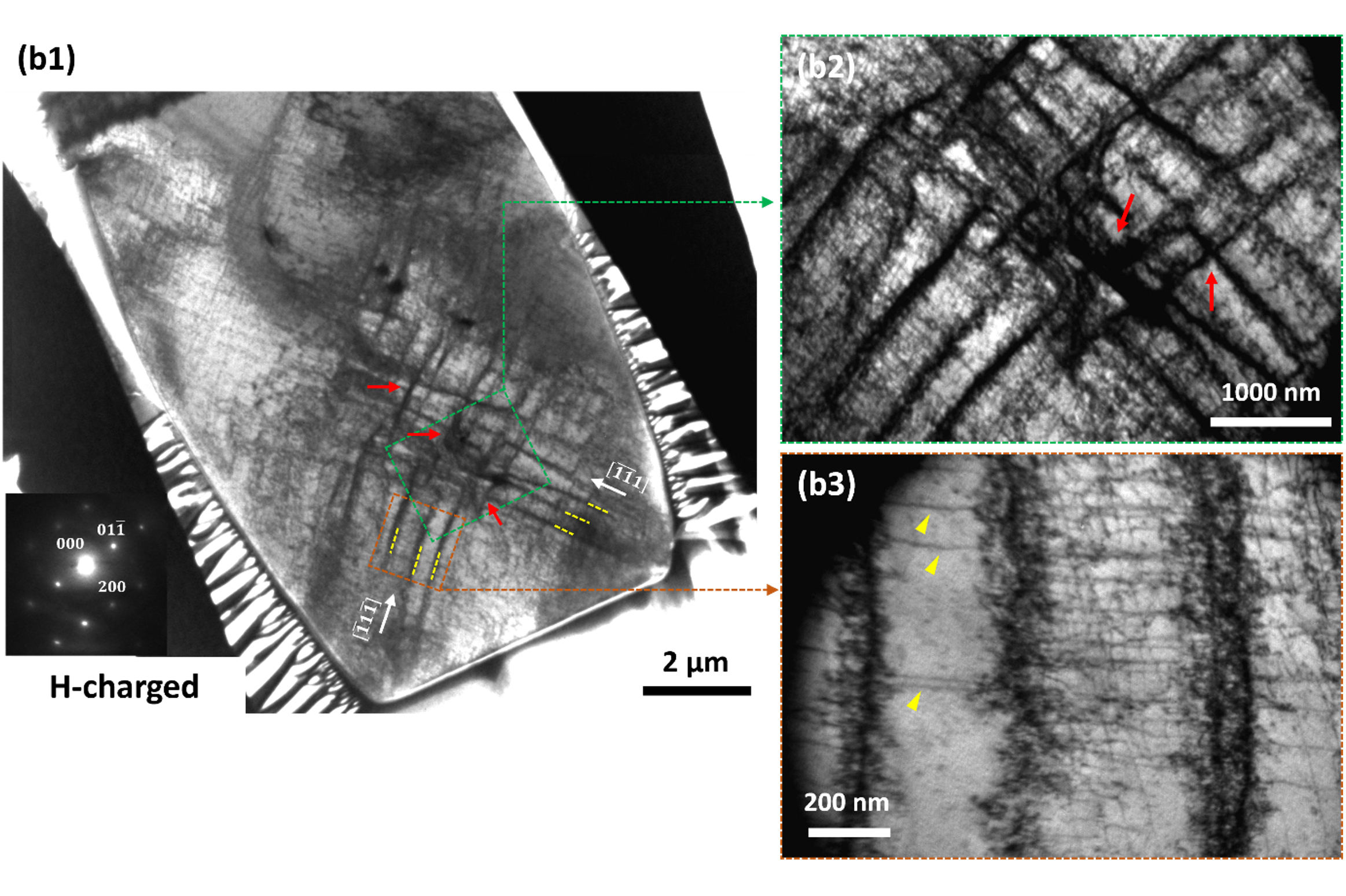}
    \caption{Bright-field TEM micrographs of the cross-sectional lamella of the deformed micropillar in: (a1-a3) H-free and (b1-b3) H-charged conditions.
    \label{fig:tem}}
\end{figure*}

H-charging was carried out in potentiostatic mode at -1.5 V (vs V$_{\mathrm{ref}}$) in a deaerated 0.1 M NaOH solution. Further details on charging optimisation are provided in the Supplementary Materials. Micropillars with a diameter of approximately 5.6 µm and a height of $\sim$15 $\mu$m were fabricated using a FEI Helios 600i focused ion beam (FIB) system on a pristine sample (30 kV, 9 nA). Micropillar compression was performed using the ASA inside a Zeiss Ultra SEM. For the compression, a doped diamond flat punch with $\sim$10 $\mu$m diameter was employed, and pillar compression was carried out in displacement-controlled mode at 50 nm/s (strain rate, $\dot{\varepsilon}\sim 3\times 10^{-3}$ s$^{-1}$). Pillars were compressed under two conditions: before starting the H-charging, i.e., H-free state and after H-charging for $\sim$2 hours, in steady-state condition (see Figure S3(b)). At least 3 pillars were compressed for each condition (see the supplementary videos). Note that pillars containing second-phase particles were avoided. The strain rate was varied by an order of 5 from $3\times 10^{-4}$ s$^{-1}$ up to 3 s$^{-1}$ for strain-rate jump testing.

Figure \ref{fig:pillars}(a) shows the engineering stress–strain curves of Fe–Cr micropillars. The loading direction is parallel to <110>, and all pillars were deformed up to $\sim$15\% strain. Yield stress is defined at 1\% plastic strain. The hydrogen-charged pillars exhibit a marginal increase in yield stress ($\sim$60 MPa) and an enhanced apparent strain-hardening rate compared to the H-free condition. In the H-free state, the stress–strain curves display multiple discrete stress drops, indicating intermittent dislocation avalanches, which is attributed to the low initial dislocation density. In contrast, the H-charged pillars deform more uniformly, with less pronounced stress drops, suggesting a more homogeneous plastic deformation and smoother dislocation movement.

As shown in post mortem SEM micrographs in Figure \ref{fig:pillars}(b), the deformation morphology of the pillar has clearly changed with the introduction of hydrogen. The H-free pillars exhibit several sharp, well-defined slip bands and shear offsets on the surface (indicated by arrows), consistent with the stress drops observed in the stress-strain curves. Furthermore, deformation appears to be dominated by localised slip events on a single family of parallel slip planes (highlighted by dotted lines). Schmid factor calculations for compression axis parallel to <011> indicate that the {112}<111> slip system has the highest resolved shear stress. The observed slip trace angles are in accordance with those estimated for the {112}<111> slip system, confirming that plastic deformation is primarily accommodated by dislocation glide on this family of planes. In contrast, the H-charged pillars show dense, shallow, and more symmetrically distributed slip traces on either side of the pillars, indicating activation of multiple non-parallel {112}<111> slip systems. The overall barrel-like appearance of the pillar further supports this observation. In addition, the spacing between slip lines is considerably smaller in H-charged pillars.

\begin{figure*}[!hb]
    \centering
    \includegraphics[width=0.99\textwidth]{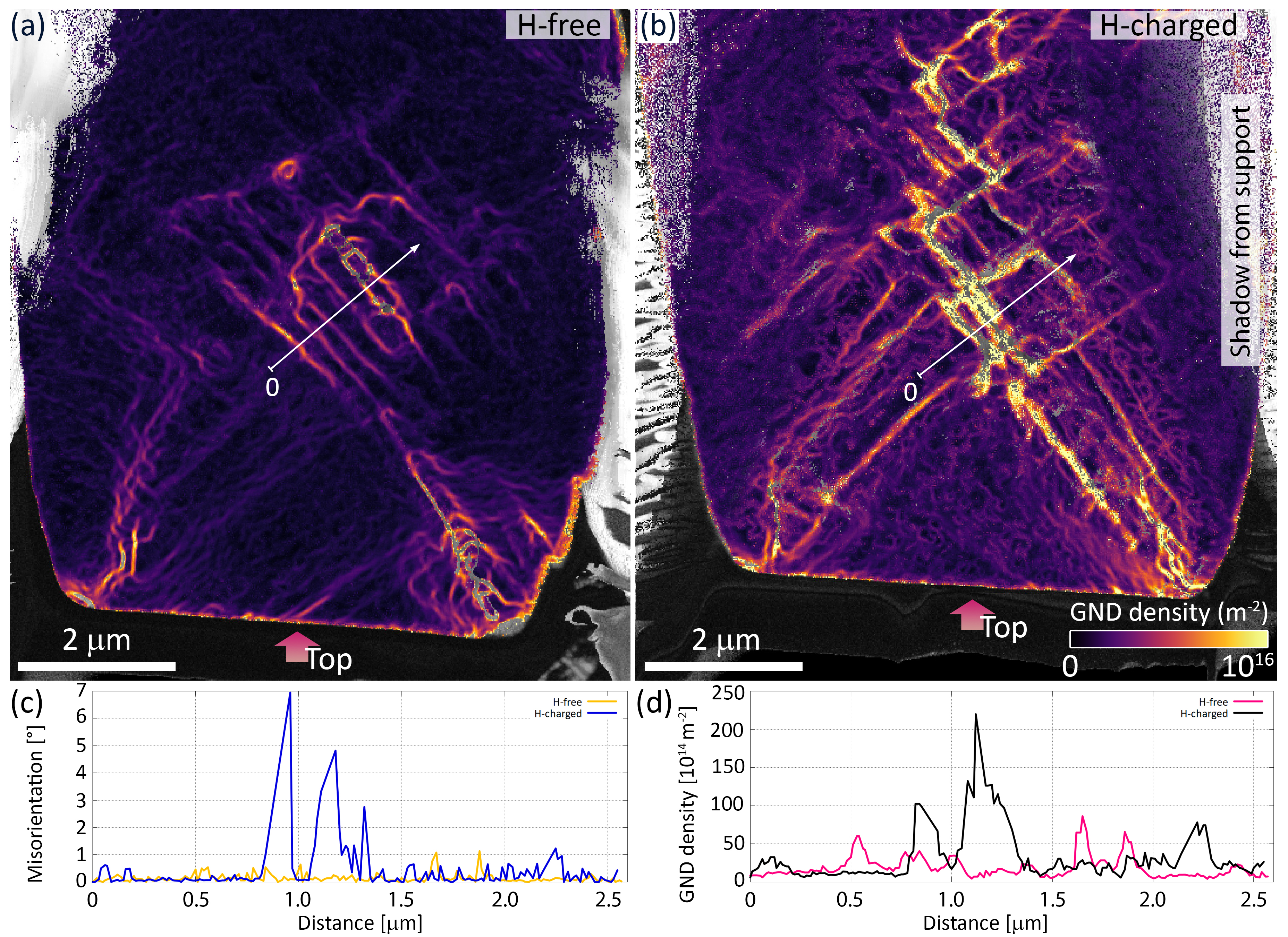}
    \caption{The GND density maps obtained from TKD analysis: (a) lamella from the H-free pillar, (b) lamella from the H-charged pillar, (c,d) comparison of misorientation angle and GND density along the line indicated in (a) and (b).
    \label{fig:tkd}}
\end{figure*}

To characterise the post-deformation dislocation microstructures, a FIB lamella was extracted from the deformed H-free-2 and H-charged-2 micropillars parallel to the loading direction. The prepared lamellae were analysed by TEM (Philips CM200 at 200 kV) and Transmission Kikuchi Diffraction (TKD, 30 kV, 2 nA) using an Oxford Instruments Symmetry 2 detector. Figure \ref{fig:tem} shows the TEM bright-field micrographs of the lamellae from H-free and H-charged pillars. In the H-free sample, multiple parallel deformation bands (dotted lines in the centre) are aligned with the surface shear offsets (indicated by arrows), consistent with the surface observations. This indicates that the dislocations have been generated at the surface and propagated into the micropillar. Similar observations were made on Fe micropillars deformed at room temperature without H \cite{Hagen.2017}. In contrast, the dislocation network in the H-charged pillar shows pronounced entanglement and junction formation in the middle of the pillar (indicated by arrows in Figure \ref{fig:tem}b) due to the interaction of dislocations gliding on non-parallel slip planes (highlighted by dotted lines). No large shear offsets were observed on the pillar surface. In both cases, the residual dislocation structure shows straight <111> type dislocation segments, suggesting that screw dislocations have controlled the plastic deformation. This is expected since edge dislocations move much faster than screw dislocations in BCC metals \cite{Itakura.2013}. Figure \ref{fig:tkd} shows the spatial variation of geometrically necessary dislocation (GND) density obtained from misorientation maps by TKD analysis of the same lamellae. Clearly, the GND density is much higher in the H-charged lamella.

\begin{figure*}[!hb]
    \centering
    \includegraphics[width=0.90\textwidth]{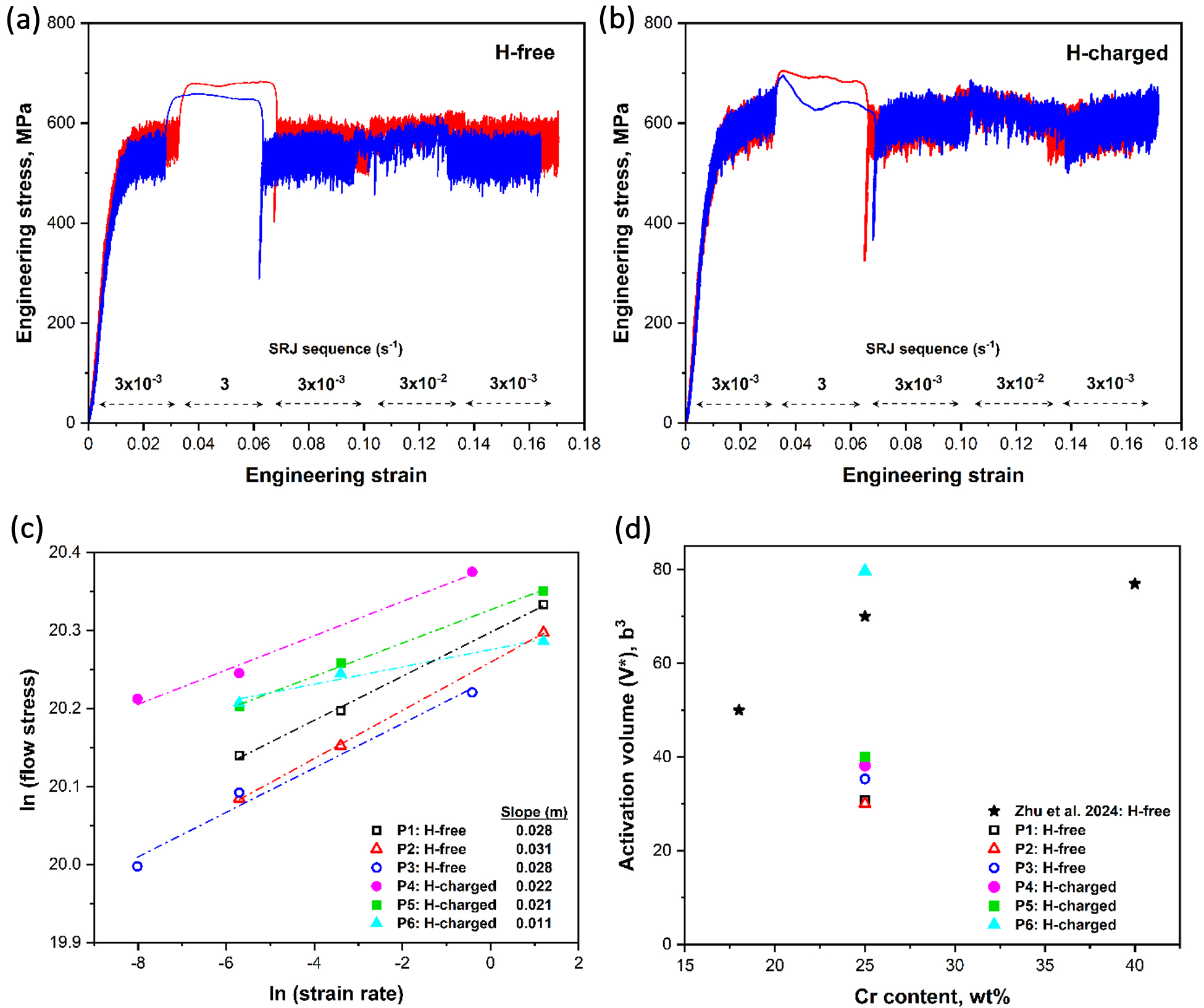}
    \caption{Engineering stress-strain curves from strain rate jump (SRJ) tests for (a) H-free and (b) H-charged micropillars. (c) ln-ln plot of average flow stress vs strain rate data (from SRJ tests). (d) Estimated activation volume for H-free and H-charged conditions, along with literature data for Fe-Cr alloys [45]. 
    \label{fig:srj}}
\end{figure*}

H-induced hardening has been reported during in situ nanoindentation in a variety of materials \cite{Stenerud.2017,Wang.2019, Lu.2019,Park.2021,Wang.2021,Soundararajan.2023,Zhao.2017} including Fe-Cr alloys \cite{Duarte.2021,Rao.2023}. The hardening was attributed to: (i) resistance to dislocation motion through increased lattice friction and solute drag effects of H \cite{Rao.2023,Stenerud.2017,Wang.2019}, and (ii) H-enhanced dislocation nucleation and multiplication, followed by entanglement leading to hardening (consistent with the defectant theory \cite{Kirchheim.2010}). In our case, the marginal increase in yield strength can be attributed to lattice friction due to H and is consistent with the estimated value of $\sim50$ MPa for Fe-21Cr reported by Rao et al. \cite{Rao.2023}. Increased strain hardening can be explained based on the H-enhanced dislocation density or solute drag effect. In BCC metals, plastic deformation is controlled by the generation and motion of kink pairs on screw dislocations \cite{Kirchheim.2012}. Solute H can affect both these processes, and the consequent softening or hardening is dictated by the interplay between the time-scales of kink-pair generation ($\tau_\mathrm{g}$) and motion ($\tau_\mathrm{m}$). According to the defectant concept, H reduces the kink-formation energy, therefore decreasing $\tau_\mathrm{g}$. Both simulations \cite{Itakura.2013} and experiments \cite{Huang.2023} support this. Thus, if the rate-limiting step is kink generation (i.e., $\tau_\mathrm{g}$ > $\tau_\mathrm{m}$), which is the case at low H concentrations (or low charging current densities), softening is expected \cite{Kirchheim.2012}. Kapci et al. have observed a reduced yield strength during micropillar compression of $\alpha$-Ti charged at 0.05 mA/cm$^2$ \cite{Kapci.2025}. On the other hand, H also has a strong tendency to segregate to the generated kinks -- which are of edge character and thus strongly attract H -- and pin them \cite{Kirchheim.2012}. Thus, at high H concentrations, the solute-drag effect becomes dominant (i.e., $\tau_\mathrm{m}$ > $\tau_\mathrm{g}$), and hardening is predicted \cite{Kirchheim.2012,Itakura.2013,Katzarov.2017,Wagih.2015,Leveau.2026}. Since the compression tests were conducted after charging the sample to a steady-state condition at a relatively high current density (2.5 mA/cm$^2$), a higher concentration of mobile H is expected in our pillars. Note that Rao et al. observed a peak hardness in the Fe-21Cr alloy under similar charging conditions \cite{Rao.2023}. The second mechanism is H-enhanced dislocation density. Solute H, by reducing dislocation formation energy, can initiate slip at a lower resolved shear stress and enhance dislocation multiplication compared to the H-free condition \cite{Kapci.2025,Chen.2013}. This leads to the activation of multiple slip systems and the eventual entanglement of dislocations gliding on intersecting slip planes \cite{Fang.2019}, resulting in forest hardening and the suppression of shear localisation. Shallow slip traces on the H-charged pillar surface support this. Similar observations were made on W \cite{Fang.2019} and $\alpha$-V \cite{Deutges.2013}. Both TEM (Figure \ref{fig:tem}) and TKD (Figure \ref{fig:tkd}) analysis clearly depict the junction formation in the middle of the H-charged pillar and show higher dislocation density than in the H-free state, supporting the above explanation. Thus, the increased yield strength and strain hardening can be attributed to the increased lattice friction, solute drag effect, activation of multiple non-parallel slip systems and enhanced dislocation density due to H.

To further understand  the underlying mechanisms, we conducted strain-rate jump (SRJ) tests under identical charging conditions. Figure \ref{fig:srj} (a,b) shows the stress-strain curves from SRJ tests (corrected for apparent strain hardening following the method outlined in \cite{Mohanty.2015}) in H-free and H-charged conditions. Figure \ref{fig:srj}(c) shows the average flow stress for each segment plotted against strain rate. The activation volume (\textit{V*}) is calculated using the following equation \cite{Zhu.2024}:

\begin{equation}
    V^*=\frac{kT}{M} \left(\frac{\partial \ln \dot{\varepsilon}}{\partial \sigma} \right),
\end{equation}

\noindent where M is the Schmid factor ($\sim 0.5$ in our case). The observed range of strain-rate sensitivity (\textit{m}) and \textit{V*} (30 -- 40 b$^3$) aligns with the literature and points towards a double-kink mechanism \cite{Zhu.2024}. The lower absolute \textit{V*} values than those reported by Zhu et al. \cite{Zhu.2024} may be attributed to microstructural differences or FIB-induced defects. In contrast to the literature data on BCC metals, in which either a decrease \cite{Wang.2013a,Wang.2013b} or no change \cite{Schafer.2025} was reported, we observed a marginal increase in \textit{V*} (except for pillar P6, which is an outlier) in the presence of H. The reduction in \textit{V*} is often attributed to the H-enhanced dislocation mobility accompanied by softening \cite{Wang.2013a,Wang.2013b,Wang.2014,Wang.2013c}. While both H-enhanced dislocation mobility (i.e., HELP) and H-solute drag decrease \textit{V*} \cite{Schafer.2025}, the former leads to softening as observed in \cite{Wang.2013a,Wang.2013b,Wang.2014}, whereas the latter results in hardening \cite{Leveau.2026,Fang.2019}. Since we observed hardening and high dislocation density (Fig. \ref{fig:tkd}) in H-charged pillars, we suggest that the marginal increase in \textit{V*} arises from two competing mechanisms: (i) H-enhanced dislocation density, which contributes to forest hardening-like effects leading to higher \textit{V*}, and (ii) solute drag on kink mobility, which reduces \textit{V*}. Together, these effects result in an activation volume intermediate between that of a pure solute-drag (low \textit{V*}) and a fully developed forest hardening regime (high \textit{V*}). Nevertheless, microscale SRJ studies using the back-side H-charging method are lacking and require further investigation to fully explain these elementary mechanisms. 

In summary, an in situ back-side H-charging setup to eliminate the charging-related surface damage effects and carry out micromechanical testing inside SEM was developed. The novel design of the setup suppresses flow vibrations and allows nanoindentation/pillar compression during H-charging, ensuring the presence of mobile H in the probing volume. In situ micropillar compression and SRJ analysis of Fe-25Cr single crystal suggest that the H results in forest hardening due to the activation of multiple slip systems and high dislocation density, and solute drag could play a definitive role in the deformation mechanisms.

\section*{Acknowledgement}
SK and LB were funded by the French National Research Agency (ANR) under the project No. ANR-22-CE08-0012-01 (INSTINCT). This article was supported by the COST (European Cooperation in Science and Technology) Action CA21121 MecaNano. The Authors would like to acknowledge the assistance of Claire Roume, Eric Garrigou, Bernard Allirand, Aman Prasad, Yoann Garnier (Mines St. Etienne), and Vincent Barnier (CNRS LGF, Mines St. Etienne) during the development phase and initial experiments. We would also like to thank the support provided by Alemnis AG (Switzerland) throughout the implementation of the experimental setup.

\section*{CRediT authorship contribution statement}
\textbf{Lavakumar Bathini:} Writing – original draft, Writing – review and editing, Conceptualisation, Methodology, Investigation, Data curation, Visualisation, Validation. \textbf{Guillaume Kermouche:} Writing – original draft, Writing – review and editing, Conceptualisation, Supervision, Resources. \textbf{Sergio Sao-Joao:} Investigation. \textbf{Fré\-dé\-ric Chris\-ti\-en:} Writing – review and editing, Supervision, Resources. \textbf{Szilvia Kalácska:} Writing – original draft, Writing – review and editing, Conceptualisation, Methodology, Investigation, Data curation, Visualisation, Validation, Supervision, Funding acquisition, Project administration, Resources.

\section*{Data Availability}
Supporting raw data are available at the Zenodo repository (\url{https://doi.org/10.5281/zenodo.21873297}) or from the Corresponding authors upon reasonable request.

\bibliography{references}

\end{document}